\documentclass{article}

\usepackage[english]{babel}
\usepackage{authblk}
\usepackage{float} 

\usepackage[letterpaper,top=2cm,bottom=2cm,left=3cm,right=3cm,marginparwidth=1.75cm]{geometry}

\usepackage{amsmath}
\usepackage{graphicx}
\usepackage[colorlinks=true, allcolors=blue]{hyperref}

\usepackage{algorithmic}
\usepackage{graphics}
\usepackage{multirow}
\usepackage{booktabs}
\usepackage{threeparttable}
\usepackage{color,colortbl}
\usepackage{amsmath}
\usepackage{makecell}

\title{Predicting the popularity of information on social platforms without underlying network structure}

\author[1,2,3,$\dagger$]{Leilei Wu}

\author[4,$\dagger$]{Lingling Yi}

\author[1,5,*]{Xiao-Long Ren}

\author[1,5,*]{Linyuan  {L\"{u}} }

\affil[1]{Yangtze Delta Region Institute (Huzhou), University of Electronic Science and Technology of China, Huzhou 313001, P. R. China}
\affil[2]{Department of Physics, University of Fribourg, CH-1700 Fribourg, Switzerland}
\affil[3]{Alibaba Business School, Hangzhou Normal University, Hangzhou 311121, P.R. China}
\affil[4]{Tencent Technology (Shenzhen) Co.Ltd, Shenzhen 518000, China}
\affil[5]{Institute of Fundamental and Frontier Sciences, University of Electronic Science and Technology of China, Chengdu 611731, P. R. China}
\affil[$\dagger$]{These authors contributed equally to this work.}
\affil[*]{Correspondence: renxiaolong@csj.uestc.edu.cn (X.-L.R.), linyuan.lv@uestc.edu.cn (L.L.)}

\begin{document}
\maketitle

\begin{abstract}
The ability to predict the size of information cascades in online social networks is crucial for various applications, including decision-making and viral marketing. However, traditional methods either rely on complicated time-varying features that are challenging to extract from multilingual and cross-platform content, or on network structures and properties that are often difficult to obtain. To address these issues, we conducted empirical research using data from two well-known social networking platforms, WeChat and Weibo. Our findings suggest that the information-cascading process is best described as an activate-decay dynamical process.
Building on these insights, we developed an Activate-Decay (AD)-based algorithm that can accurately predict the long-term popularity of online content based solely on its early repost amount. We tested our algorithm using data from WeChat and Weibo, demonstrating that we could fit the evolution trend of content propagation and predict the longer-term dynamics of message forwarding from earlier data. 
We also discovered a close correlation between the peak forwarding amount of information and the total amount of dissemination. Finding the peak of the amount of information dissemination can significantly improve the prediction accuracy of our model. Our method also outperformed existing baseline methods for predicting the popularity of information.
\end{abstract}

diffusion, online social network

\section{Introduction}\label{Introduction}

With the booming development of communication technologies and mobile services, online social networks enable billions of users to create and share information worldwide freely. Reading and reposting online content has become a significant way for individuals to communicate and express opinions \cite{brady2021social,zhao2010weak}. To this end, the dissemination of information plays a fundamental role in our daily life and is of great economic value and practical significance \cite{Lazer2009,freelon2020false}.
The capacity to collect, clean, and analyze large-scale data has transformed the field of social network analysis and empowers scientists with enhanced convenience and efficacy in conducting large-scale study~\cite{Wasserman1994,Aggarwal2011,Pastor2015,Dirk2013}.
The study of information spreading in social networks has become one of the core topics in computational social science~\cite{Lazer2009,Giles2012,conte2012manifesto} and network science~\cite{barabasi2016,Newman2010}. It attracts increasing attention from fields such as sociology, physics, computer science, \textit{etc}.

Among the above, the popularity prediction of information on social platforms is a crucial issue that has been widely concerned by both academic and industrial researchers in recent years \cite{szabo2010predicting,cheng2014can,liao2019popularity,chen2019information,zhou2021survey,yu2015multi,yu2017tiirec,2022PNASPredicting}. By ``popularity", we usually mean the final amount of viewing, collecting, forwarding, or sharing of information in networks~\cite{szabo2010predicting}, depending on the actual setting of each research. 

Firstly, let us briefly review the research progress on the popularity prediction of information.
As one of the most classic studies, Szabo and Huberman \cite{szabo2010predicting} analyzed the popularity of content submitted to Digg and YouTube, where popularity means the number of votes on Digg or the number of views on YouTube, respectively. A strong linear correlation was discovered between the logarithmically transformed popularity of content in early and later time periods. The authors proposed a log-linear model-based Linear Regression (LR) method to predict the popularity. See more details of this method in Section \ref{BaselineAlgorihtm}.  

Inspired by the above approach, the linear regression with degree model (LR-D) \cite{zhao2015seismic} was proposed to predict the popularity of the information in a greater variety of data sets by considering the cumulative degree of the users who reshare content.
Further, Bao \textit{et al.}~\cite{Bao2013} found a close relationship between the popularity of the information and the structural diversity of the social network. Specifically, there exists a strong negative/positive near-linear correlation between the final popularity and its link density/diffusion depth over time. Thus, the final popularity of information can be computed by linear regression with the structural characteristics (LR-S) model.

From another viewpoint, a user who has forwarded a message may trigger another user to forward the message with a probability. 
By considering the underlying arrival process of information, and the aging effect and reinforcement effect in the spreading process, Gao \textit{et al.} \cite{Gao2015} proposed a model, named Exponential reinforcement and Time Mapping process (PETM), which combines the reinforced Poisson process model with a power-law relaxation.
Based on the theory of self-exciting point processes, Zhao et al. \cite{zhao2015seismic} developed a Self-Exciting Model of Information Cascades (SEIMIC) method to predict the future sharing volumes of given posts on Twitter. The SEISMIC only requires the timestamps of reposts and the number of followers of the users.

From the empirical analysis, it is easy to find that a handful of vital users \cite{Lu2016} dominate the spreading of information on social networks. Taking into account this phenomenon, Gao \textit{et al.} \cite{Gao2016} propose a mixture process to predict the popularity of information. 

Besides the above algorithms, an enormous amount of research has been conducted to predict the popularity of information on social networks recently \cite{yu2020prediction, wu2019smp,  zhang2022anytime, wang2021incremental,chen2022graph}. 
These research advances shed light on the applications spanning from communication, decision-making, cooperation, viral marketing, and advertising to prompt user-generated content such as blogs and scientific papers and understanding the evolution of information cascades online.

However, these methods either rely heavily on complicated features that are time-varying and cannot be easily extracted from multilingual and cross-platform content, or on the underlying network structures or properties that are often difficult to obtain. 
In this article, we analyzed several empirical data sets and found that the information-cascading process is best characterized as an activate-decay dynamical process. Based on our findings, we propose an Activate-Decay (AD)-based algorithm for predicting the long-term popularity of online content solely based on their early repost amount, without requiring knowledge of the social network structure or content properties.
The results show that our method uses the forwarding amount of information in WeChat within the first two hours to forecast its popularity for seven days with remarkable accuracy.
Furthermore, we identified a close correlation between the peak of the amount of information dissemination and the total amount of dissemination. As long as the peak of the amount of information dissemination can be found, the prediction accuracy will be significantly improved. 
Our method also outperformed existing baseline methods for predicting the popularity of information.

Following the above brief introduction to the problem we are investigating, the rest of this paper is structured as follows. Firstly, we conduct empirical analyses of two data sets of information-forwarding processes across the Weibo and WeChat platforms. Our analysis describes the rise and fall of information as an active-decay dynamical process, which provides insight into attempts to model and predict information transmission. Secondly, we propose a model based on the (Bi)Hill equation from biochemistry, which has limited parameters and can predict the popularity of information without requiring knowledge of the underlying structure of social networks or content features. Finally, we perform experiments to demonstrate the effectiveness of our proposed method.

\section{Materials and Methods}
In this section, we begin by presenting an empirical analysis of two prominent social network platforms: WeChat and Weibo. We then use the observed spreading patterns of information on these platforms to develop a dynamic process that describes the rise and fall of information over time. Using this proposed dynamic process, we are able to predict the popularity of information.

 \subsection{Empirical data analysis}
To begin with, let us provide a brief introduction of the data sets utilized. 

The WeChat data set comprises over 90,000 news articles, including political news, economic news, legal news, military news, scientific and technological news, cultural and educational news, sports news, social news, etc., and their forwarding records between the individuals in the WeChat social platform from June 2nd to June 8th, 2016, was created in a collaboration project with Tencent's WeChat department. 
The forwarding records were collected from individuals sharing in timelines, group chat, and  individual forwarding. The data includes the message $id$ and the time $t$ when a message is forwarded. The forwarding records of all messages in this data set were anonymized. 

The Weibo data set, obtained from a competition hosted by Wolong Big Data on DataCastle (\url{https://challenge.datacastle.cn}), consists of roughly 30,000 microblogs, with over 17,840,000 forwarding records. Weibo is commonly referred to as the "Twitter of China." The messages in the Weibo data set are mainly short paragraphs with at most 140 Chinese characters, with or without pictures. The data set includes the content of microblogs, the users who published or forwarded the microblogs, the publish and forward time $t$, and the following relationship between users. In this research, we only use the ids and publish/forward times of the microblogs.

\begin{figure}[H]
  \centering
  \includegraphics[width=13cm,height=13cm]{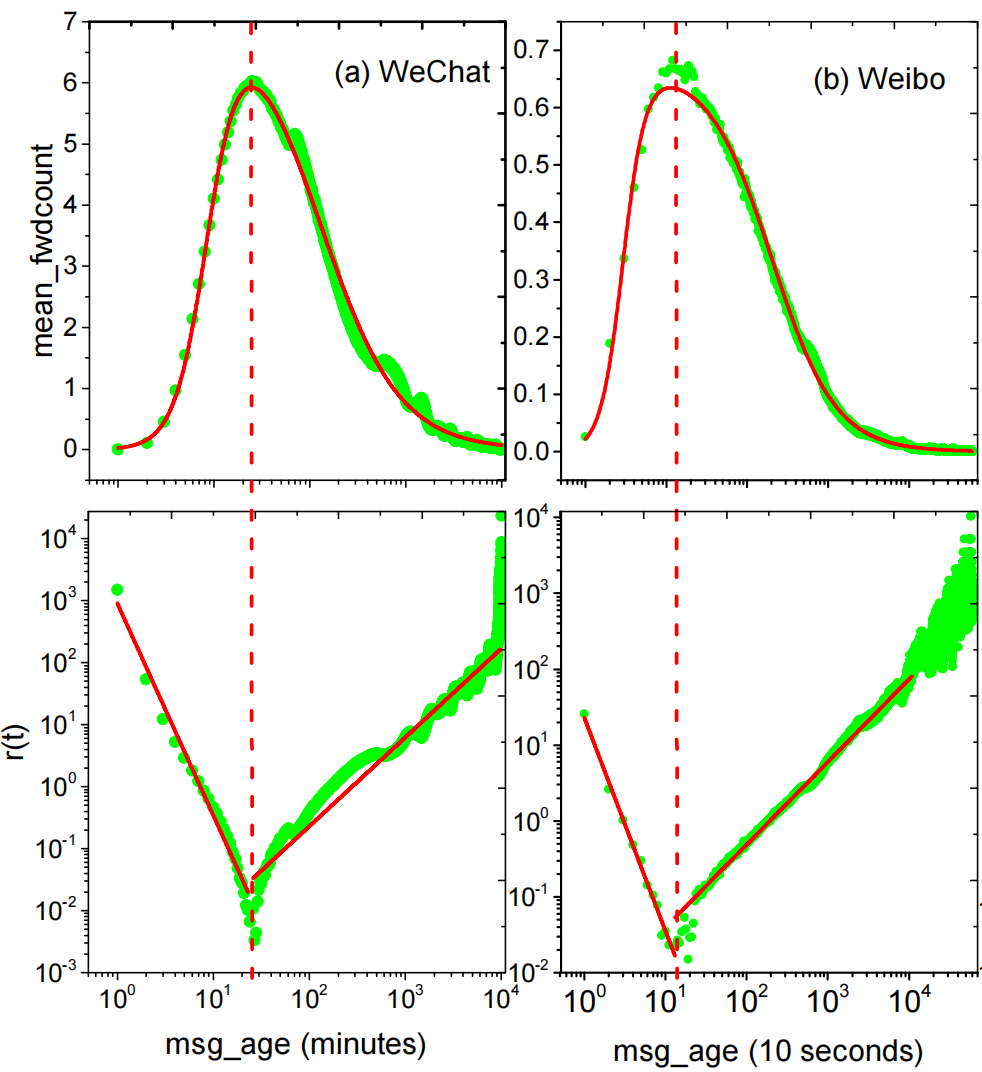}
 \caption{ The average forwarding amounts of information on WeChat and Weibo display similar statistical trends over time.
 In this figure, the upper row depicts the relationship between the average forwarding amount and time unit, with the horizontal axis scaled to (a) 1 minute and (b) 10 seconds for WeChat and Weibo, respectively.
 The lower row is the trend of the average forwarding volume from its peak value over time. In terms of time, it takes time for the amount of news dissemination to reach the average peak, and the dissemination of information on different social platforms shows a large gap in the rate of information dissemination. Obviously, the transmission rate of information on Weibo is faster than on WeChat. On average, for WeChat, it takes less than 30 minutes (1800 seconds) for a message to reach its peak from generation to transmission per unit time, while it takes only 200 seconds for Weibo.}
\label{fig2}
\end{figure}

In order to better analyze the collective forwarding pattern of different messages, we standardize the timestamp of all the forwarding records in the two data sets and note the time when the message was released as $t=0$. In Figure \ref{fig2}, we show the average amounts of information forwarded on WeChat and Weibo
exhibit varying statistical trends over time. The figure's top row depicts the correlation between the average forwarding amount and time unit. For WeChat and Weibo the horizontal axis scale is (a) 1 minute, (b) 10 seconds.
The X-axis is logarithmic. On average, it takes less than 30 minutes (1800 seconds) for a message to reach its peak from generation to transmission per unit time, while it takes only 200 seconds for Weibo.  
After passing the peak period, the forwarding volume of all messages gradually decreases over time. Figure \ref{fig2} indicates that the entire process can be divided into two stages, namely active and decay. The active stage is very fast to reach the peak point while the decay stage lasts a very long time. To gain a comprehensive understanding of the entire process, the x-axis of Figure \ref{fig2}'s top row was plotted using a logarithmic scale.
To visually show the rate of change of the forwarding number before and after reaching the maximum value, i.e., the maximum forwarding volume per unit time, after the information was released, the lower row of Figure \ref{fig2} was plotted in a log-log coordinate. The shapes of the curves indicate that the change in the information's dissemination rates roughly follows a power law.
The dissemination of news takes a little time to reach the average peak, and the rate of information dissemination on different social platforms exhibits a subtle difference. Notably, Weibo shows faster transmission rates than WeChat. Please find more analysis in Section \ref{ADModel}.

In this research, our goal is to predict the final number of forwarding of a given message. 
Building on the empirical analysis mentioned above, we formulate a mathematical method that captures the rise and fall of the information dissemination process depicted in Figure \ref{fig2}. Our model enables us to predict the future shares of a piece of information by examining its sharing history, indicating whether the sharing cascade has undergone an initial stage of rapid expansion and identifying the messages that are most likely to be shared extensively in the future. 
After clearing and filtering the records, the data sets were divided as a train set and a test with 75\% and 25\% of the messages according to their real release time.

\subsection{The Activation-Decay Model} 
    \subsubsection{The Hill equation and BiHill equation} 

The Hill equation, which was introduced by A.V. Hill in 1910 \cite{Hill1910}, is a biochemical characterization equation that has been widely utilized for analyzing nonlinear quantitative drug-receptor relationships \cite{goutelle2008hill}. Additionally, the Hill equation and its variant BiHill can also be used to describe the nonlinear transmission mathematically \cite{frank2013input}. 
Hill equation can be expressed as follows \cite{nelson2013lehninger}
 \begin{equation}
\theta  = \frac{1}{1 + (\frac{K_A}{\lvert L \rvert})^n},
\end{equation}
$\theta$ is the fraction of occupied sites where the ligand can bind to the active site of the receptor protein. $\lvert L \rvert$ is free (unbound) ligand concentration. $n$ is the Hill coefficient, which describes the synergy and is a measure of super sensitivity (i.e. the steepness of the response curve). Generally speaking, $n$ determines the cooperativity of ligand binding in the following way: $n>1$, positively cooperative binding: Once a ligand molecule is bound to the enzyme. the affinity of the enzyme for other ligands will increase. $n<1$, negatively cooperative binding: Once one ligand molecule is bound to the enzyme, its affinity for other ligand molecules decreases. $n=1$, noncooperative (completely independent) binding: The affinity of an enzyme for a ligand molecule does not depend on whether a ligand molecule has been bound to it. 

 We apply the Hill function to the process of information propagation, take it as the function of time $t$, and its equation form is expressed in the following:
 \begin{equation}
Hill(t) = \frac{p}{1 + (\frac{k}{ t })^{h}},
\end{equation}
where $p>0$, $k >0$, $ h >0$, are three parameters. And when $h >0$, the system is in the activation effect, and the curve rises; when the $ h >0$, the system is in the inhibition effect, and the curve decays. 

 The Biphasic Hill equation, abbreviated as the BiHill equation, indicates that activation and inhibition exist in the whole system at the same time. The BiHill equation is expressed as follows:
\begin{equation}
BiHill (t) = \frac{P_m}{[1 + (\frac{K_a}{ t })^{H_a}] * [1 + (\frac{K_i}{ t })^{H_i}]},
\end{equation}
where $p_{m}>0$, $K_a>0$, $K_i>0$, $H_a>0$, $H_i>0$ are  the maximum value, the half-maximal activating value, the half-maximal inhibitory value,  the activation Hill coefficient, the inhibitory Hill coefficient of $BiHill (t)$, respectively. See the upper row of Figure \ref{fig2}. Applying this function to the process of information dissemination, the effects of activation and inhibition mechanisms in information dissemination are consistent with the mathematical meaning of this formula. 

   \subsubsection{The Activation-Decay Model.} \label{ADModel}
 According to the empirical analysis, the average forwarding amount of messages changes over time. In the beginning, the average amount of forwarding in unit time increases fast. However, when reaching the peak, i.e., the maximal amount, it decays slowly, until close to $0$. Define an index $r(t)$ to measure the degree of information dissemination approaching the peak value,
\begin{equation}
r(t)  = \frac{q_{max} - q(t)}{q(t)},
\label{rt_c}
\end{equation}
 where $q_{max} = max[q(t)]$. It then clearly follows that:
 \begin{equation}
r(t)  = K * t^H,
\label{rt_f}
\end{equation}
where $K$ and $H$ are two parameters. It is deduced that
\begin{equation}
q(t)  = \frac{q_{max}}{1 + K * t^H}.
\end{equation}
Obviously, it is just a form of the Hill equation. $r(t)$ is a quantitative index, and the greater its value is, the closer the amount of propagation per unit granularity is to the peak value. We have verified that $r(t)$ is a segmented function with a log-log presenting the shape of "V" according to the real social network data. When $H<0$, $r(t)$ is the "V" decaying part in the double log coordinate, while $H > 0$, it is the "V" rising part. See Figure \ref{fig2}.

In the process of disseminating information to a broader audience, there are often two opposing forces at play: activation and decay. Activation refers to factors that contribute to the spread or promotion of information, while decay refers to factors that inhibit or slow the spread of information. These two forces interact with each other in a dynamic and game-like manner, influencing the ultimate outcome of the information dissemination process. This interaction between activation and decay factors can be complex and multifaceted, as various factors may contribute to the spread or inhibition of information. 

We consider that the process of information dissemination is the interaction of activation and decay factors, and a game exists between them. Before the peak value of propagation per unit granularity, the activation state plays a leading role. After the peak value, the decay factor begins to dominate. Hence $q (t)$  will show a process of rising and then decaying over time. Therefore, we define 
\begin{equation}
F  = \frac{1}{1 + K * t^H}.
\end{equation}
When $H < 0$, $F$ is the motivation factor, and when $H > 0$, $F$ is the decay factor. 

Based on the analysis above, and the random fluctuations can be regarded as an additive noise term, we construct a prediction function named AD function, $q(t) = \alpha * q_{max} * Activation factor * Decay factor + Error function$. That is,
\begin{equation}
q(t) = \alpha * q_{max} * \frac{1}{1 + K_a * t^{H_a}} * \frac{1}{1 + K_d * t^{H_d}} + e^\beta.
\end{equation}
Where $\alpha$  and $\beta$ are harmonic parameters, which acquire from the historical date training. Also, it can be shown as: 
\begin{equation}
q(t) = \alpha * BiHill(t) +   e^\beta.
\label{BiHill}
\end{equation}
Therefore, we can directly use the Bihill equation in OriginLab to fit the parameters $K_a$, $K_d$, $H_a$, $H_d$. 

From the calculation of the average propagation of all messages selected by the system to the forwarding of each message, the prediction function is:
\begin{equation}
Q(t) = \alpha * Q_{max} * \frac{1}{1 + K_a * t^{H_a}} * \frac{1}{1 + K_d * t^{H_d}} + e^\beta.
\end{equation}
Where $Q_{max}=max[Q(t)] |_0^{(T_{known})}$. Then the propagation total amount of each message in $T$ days is 
\begin{equation}
{Q_{id}}^T = \sum_{0}^{T_{known}} Q(t)_{id} + \sum_{T_{known}+1}^{T} Q(t)_{id}.
\label{total_amount}
\end{equation}
Except for $Q_{max}$, other parameters can be obtained from historical data training. That is, we only need to know the peak value of information dissemination, and we can predict the information dissemination. In fact, the amount of social network information dissemination will reach its peak in a short time, with WeChat within 30 minutes and Weibo within 5 minutes, see Figure \ref{fig2}.

\subsubsection{The Algorithm for Popularity Prediction based on Activation-Decay Model.}
Assume that we have propagation data of $N$ messages in $T_{known}$, to predict the total information propagation ($T > T_{known}$) after $T$ time, note $id$ is the message, the amount of $id$'s being forwarded at t is $Q(t)_{id}$ and the average amount of $N$  messages is $q(t)$:

\begin{figure}[H]
  \centering
  \includegraphics[width=12cm]{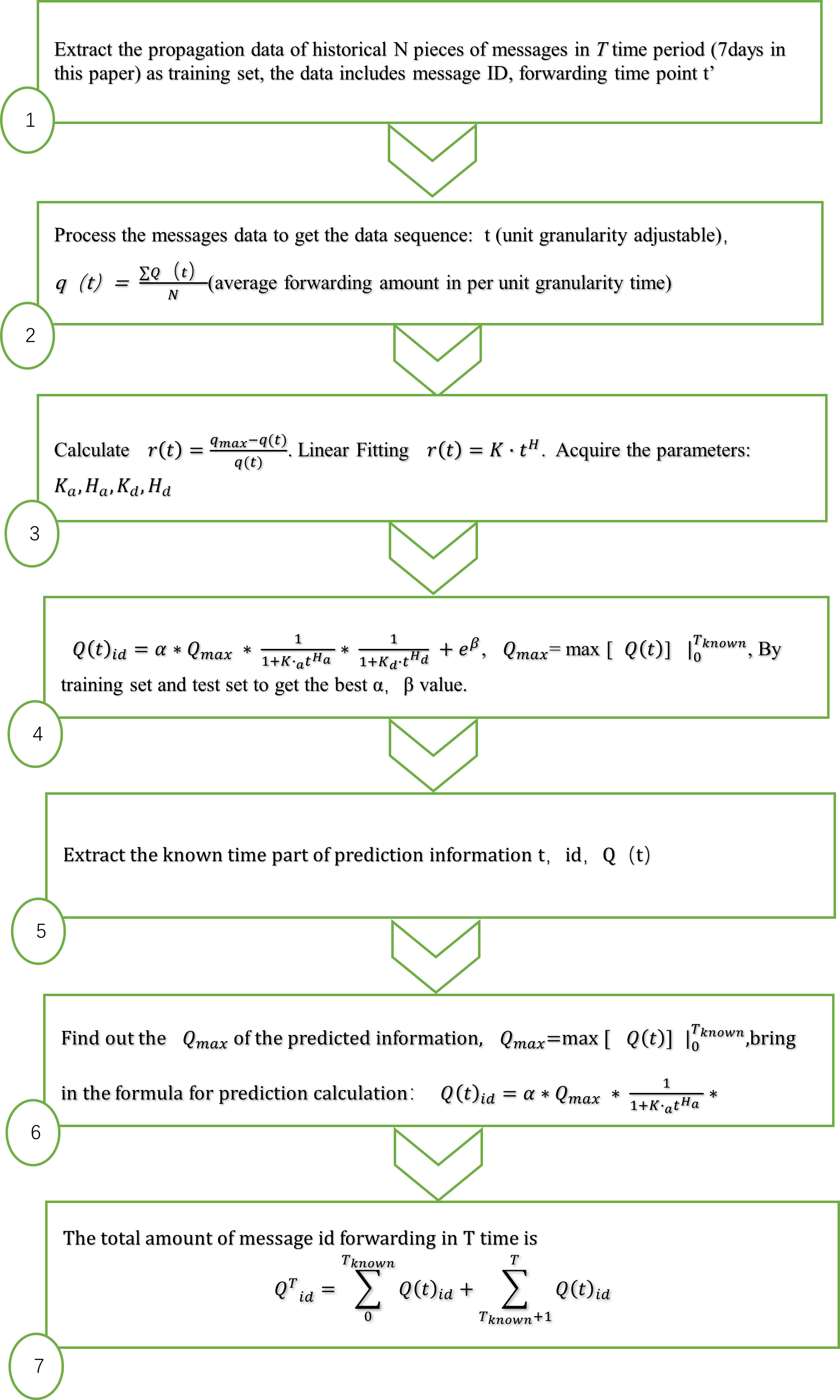}
  \caption{The flow chart of the proposed AD algorithm.}
  \label{fig_flowchart}
\end{figure}

\textbf{STEP 1} Gaining model parameters from historical data sets,$ K_a, H_a, K_d, H_d$, as shown in Figure \ref{fig_flowchart}. \textcircled{1} - \textcircled{3}:

     $(1)$. Taking the time of each message generation as the zero time, obtain the forward amount in every unit time (unit granularity adjustable). Process $N$ messages' forward amount  in $T$ period into data sequence, $t$, $id$, $Q(t)_{id}$. 
     
     $(2)$. Calculate the average amount of these $N$ messages in $T$ period time $q(t)=\frac{\sum{Q(t)_{id}}}{N}$, which yields date sequence $t$, $q(t)$.
     
     $(3)$. Estimate the parameters $ K_a, H_a, K_d, H_d$ from Eq.\eqref{rt_c} and Eq.\eqref{rt_f}, or directly obtain these parameters by fitting through BiHill equation from Eq. \eqref{BiHill}, see Figure \ref{fig2}.

\textbf{STEP 2 } Obtaining best parameters, $\alpha$ and $\beta$, by training set and test set, as shown in Figure \ref{fig_flowchart}. \textcircled{4}.

     $(1)$. The training set data is divided into two parts with the known maximum time $T_{known}$ (which can be set by oneself): the $0-T_{known}$ part is the known information set, and the $T_{known}-T$ part is the information set for prediction. e.g., if the information propagation data of 10 minutes is known, i.e., the data within $0-10$ minutes are available, and the rest is a test set.
     
     $(2)$. Find out the $Q_{max} = max[Q(t)]|_0^{T_{known}}$, calculate  the total propagation amount of each message from Eq. \eqref{total_amount}. The calculated value of the propagation amount of each message is compared with the actual propagation amount and calculate the average absolute error $MPAE$. When $MAPE$ is minimum, the parameters $\alpha$ and $\beta$ are the optimal parameters. 

\textbf{STEP 3 } Put the Related parameters ($\alpha, \beta, K_a,H_a, K_d, H_d$) into the AD algorithm to predict the propagation quantity of the information to be predicted, as shown in Figure \ref{fig_flowchart}. \textcircled{5}-\textcircled{7}.

\subsection{Evaluation Metrics for the Prediction Algorithm}

In this subsection, the evaluation metrics of the prediction algorithms used were introduced briefly.  

  \subsubsection{APE and MAPE}
    
    APE (Absolute Percent Error) is used to measure the relative error between the predicted value and the real value on the experimental data set. APE is defined as:
\begin{equation}
APE =  \frac{|Q_{id}^{predicted} - Q_{id}^{real}|}{Q_{id}^{real}} * 100 \%.
\label{total_APE}
\end{equation}
The lower the value of APE, the better the accuracy of the prediction model.

    MAPE (Mean Absolute percent error) is the average value of APE in the system, which is used to measure the relative errors between the average predicted value and the real value on the test set. MAPE is defined as:
\begin{equation}
MAPE = \frac{1}{N} * \sum_{1}^{N} \frac{|Q_{id}^{predicted} - Q_{id}^{real}|}{Q_{id}^{real}} * 100 \%.
\label{total_MAPE}
\end{equation}
Also, the lower the value of MAPE, the better the accuracy of the prediction model.
    
    \subsubsection{TIC}
    The TIC (Theil inequality coefficient) is an indicator to measure the prediction ability of the model. The smaller the general value is, the better the prediction ability of the model is. The TIC is defined as:
\begin{equation}
TIC =\frac{\sqrt{\frac{1}{N}*\sum_1^{N}(Q_{id}^{predicted})^2}}{\sqrt{\frac{1}{N}*\sum_1^{N}(Q_{id}^{predicted})^2} + \sqrt{\frac{1}{N}*\sum_1^{N}(Q_{id}^{real})^2}}.
\label{total_TIC}
\end{equation}

Therefore, the value range of this coefficient is 0-1. The closer it is to 0, the smaller the root mean square of unit error, that is, the closer the predicted value is to the actual value, the better the model fitting effect will be.

\subsection{Baseline Algorithm}\label{BaselineAlgorihtm}
As discussed in the Introduction section, there are currently numerous ways to predict popularity, including three main categories. These are predictions of early popularity \cite{szabo2010predicting}, influence factors \cite{he2014predicting,bandari2012pulse}, and cascade propagation \cite{Bao2013,kupavskii2012prediction,li2013popularity}. To validate the accuracy of our prediction method, we chose a typical popularity prediction algorithm \cite{szabo2010predicting} as the baseline method. The authors performed a logarithmic transformation on the popularity of submissions of online content from two content-sharing portals, Youtube and Digg. They found a strong correlation between the early and later times and used this relationship to predict the future popularity of messages. 
\begin{equation}
\ln{N_{s} (t_2)} = \ln{N_{s} (t_1)} + \sum_{\tau = t_1}^{t_2} \eta (\tau),  
\label{Baseline}
\end{equation}
where $N_s(t)$ is the popularity of message $s$ at time $t$, $t_1$ and $t_2$ are two arbitrarily points in time,
$t2 > t1$, and $\eta (\tau)$ refers to independent values drawn from a fixed probability distribution.

\section{Experimental Results}

The performance of the prediction model will be shown in this section. We apply three error function indicators: APE, MAPE, and TIC. We evaluate both the AD algorithm and the Baseline algorithm for data on WeChat and Weibo, by comparing the performance of the MAPE, TIC, and APE.

\begin{figure}[H] 
\centering
\begin{tabular}{ccc}
  \includegraphics[scale = 0.3]{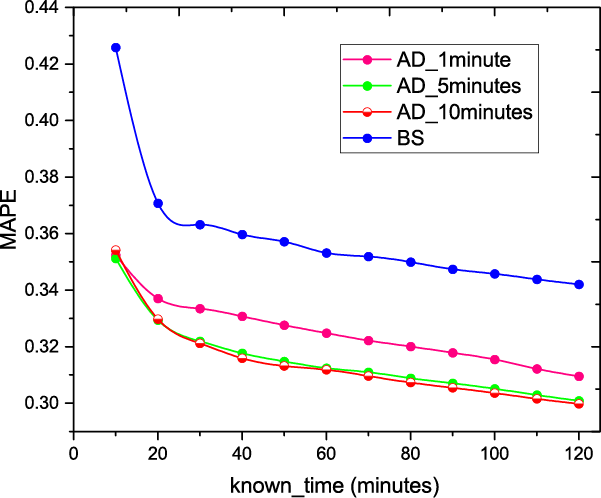}
  &\includegraphics[scale = 0.3]{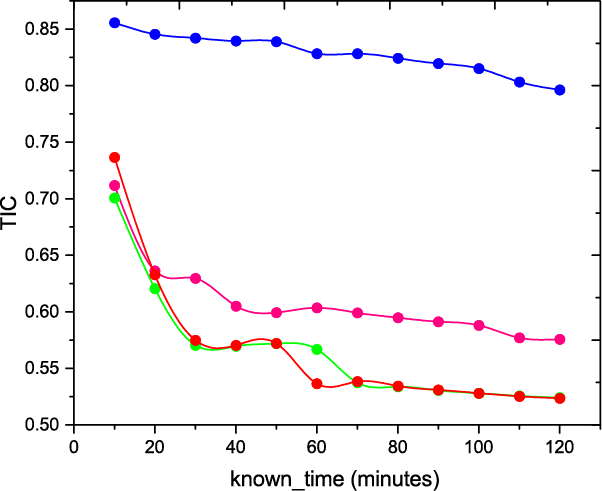}
  \\
 \includegraphics[scale = 0.4]{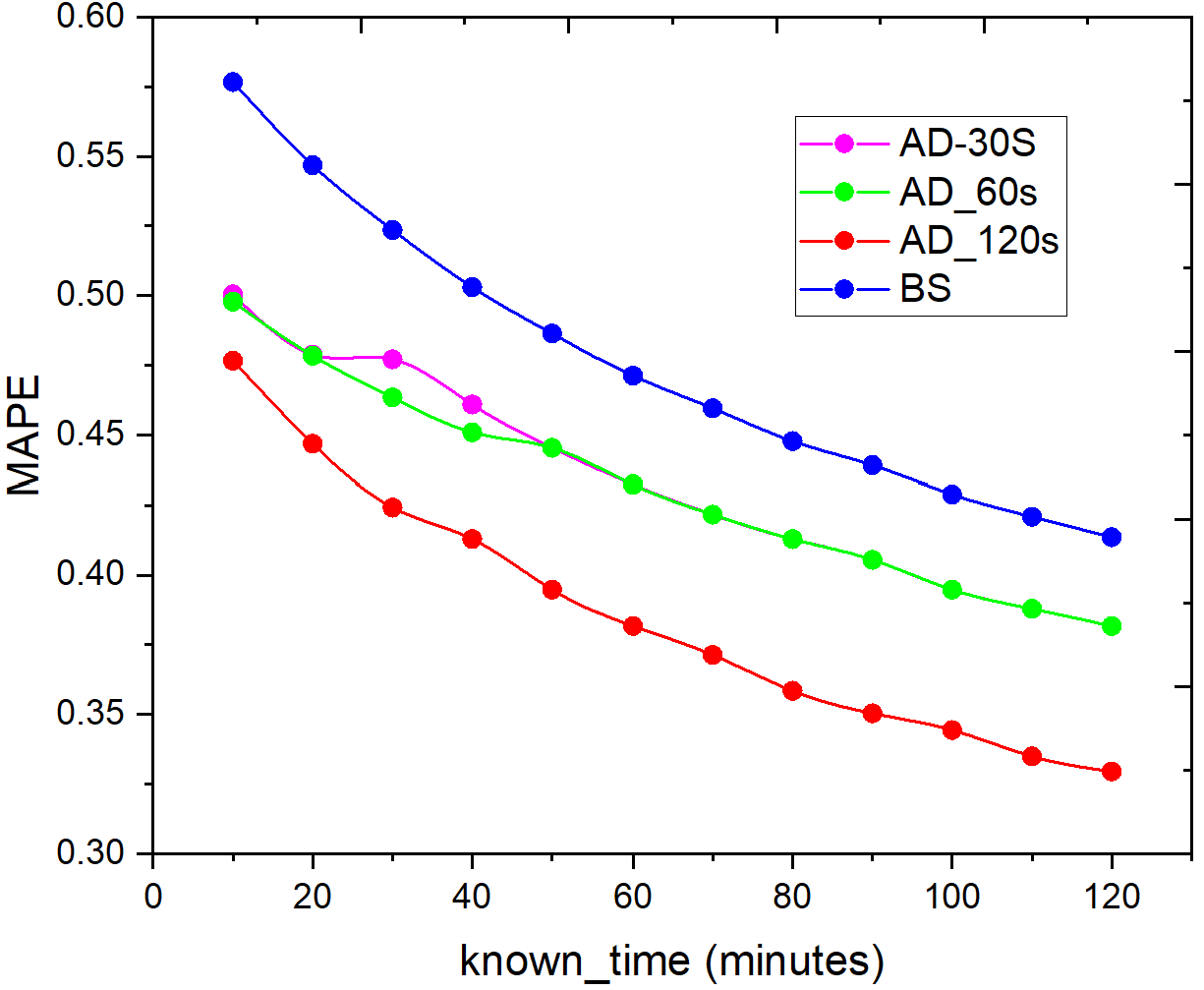}
  &\includegraphics[scale = 0.4]{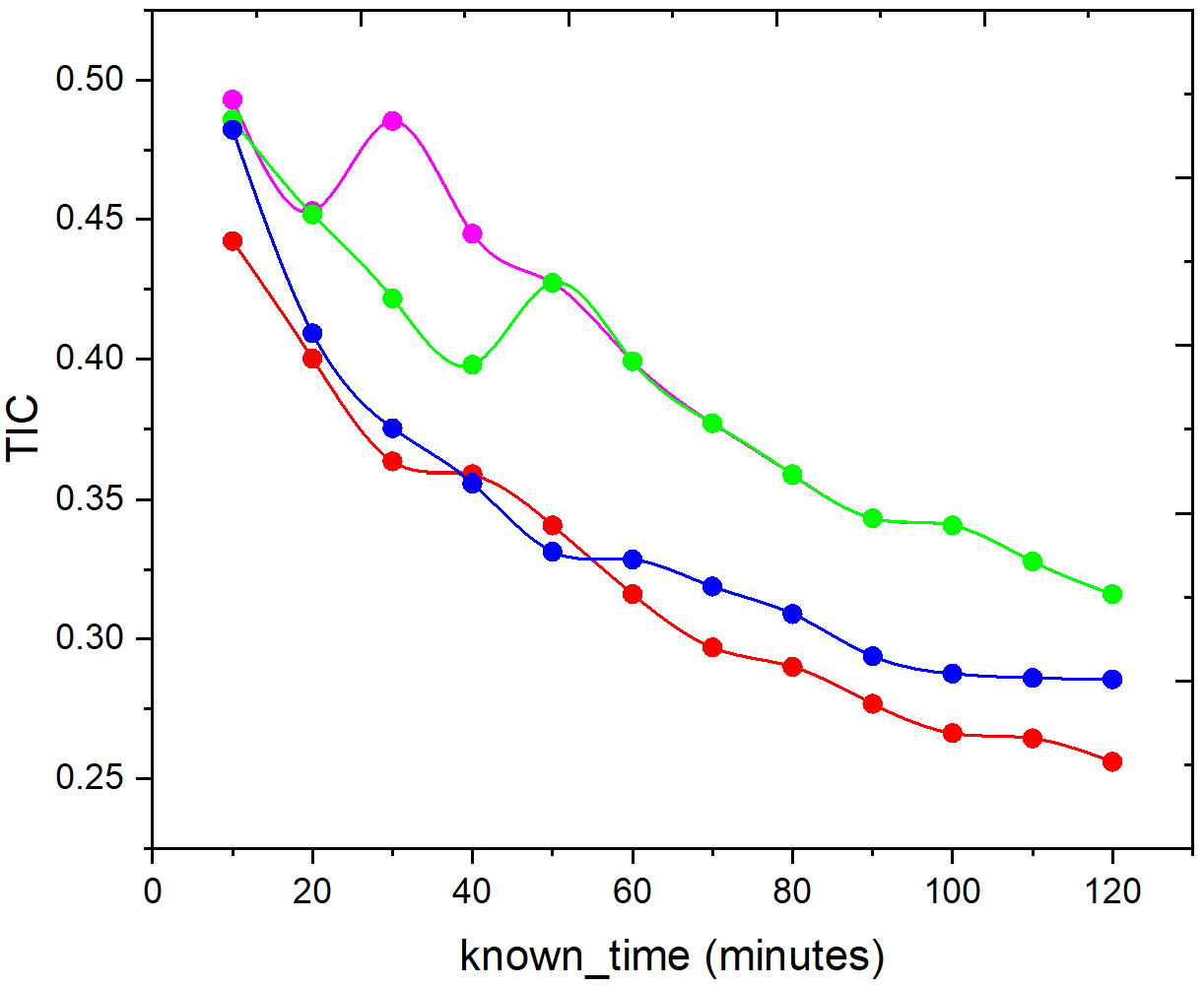}
  \\
  (a) MAPE index   & (b) TIC index
\end{tabular}
\caption{Predicting the final forward amount of messages after seven days on the basis of knowing $T_{known}$ period of information. The upper row of the figure is the results on the WeChat data set, while the lower is on the Weibo data set. The X-axis represents the known propagation time. The Y-axis means that the prediction accuracy varies with the time of known information transmission. The granularity of extracted data would affect the accuracy of AD algorithm prediction. In the upper part (WeChat) of the figure, the prediction result would reach a relatively optimal level when the unit time was 10 minutes, while in the lower part (Weibo) of the figure, the unit time was 120 seconds. These results indicate that the proposed AD algorithm outperforms the baseline (BS) algorithm.}
\label{7days_MAPE}
\end{figure}

 \subsection{Prediction of the Popularity of Information}

In Figure \ref{7days_MAPE}, we compare the performance of the AD algorithm and Baseline algorithm(called BS algorithm) on WeChat (with message number $N=31247$) and Weibo (with message number $N=25467$) social networks. We can draw the following conclusions: $(1)$AD algorithm: Within a certain granularity range, as the granularity becomes larger, the accuracy will increase, but it will not continue to improve as the granularity increases. It can be seen from the figure that the optimal value on WeChat data is obtained when the granularity is 5 minutes, and the better value on Weibo is 120 seconds.
$(2)$ With the growth of the known information time series($T_{known}$), the effects of the two algorithm 
are getting better and better. In WeChat data, the AD algorithm outperforms the baseline  algorithm(BS), no matter in MAPE or TIC index. In the Weibo data, the AD performed better than the BS at any granularity in the MAPE index. For TIC indexes, the AD algorithm does not perform better than the BS algorithm when the granularity is 30 seconds or 60 seconds. However, the AD algorithm begins to show its advantages when the granularity is 120 seconds.
$(3)$ After the granularity is changed, with the increase in the known propagation time, the accuracy rate is better, the reason should be that the peak value of some information may appear over a long time. If the time is short, the true peak of the information has not yet appeared when the statistics are calculated, which affects the accuracy.

\begin{figure*}[htb]
\centering
\begin{tabular}{ccc}
  \includegraphics[scale = 0.3]{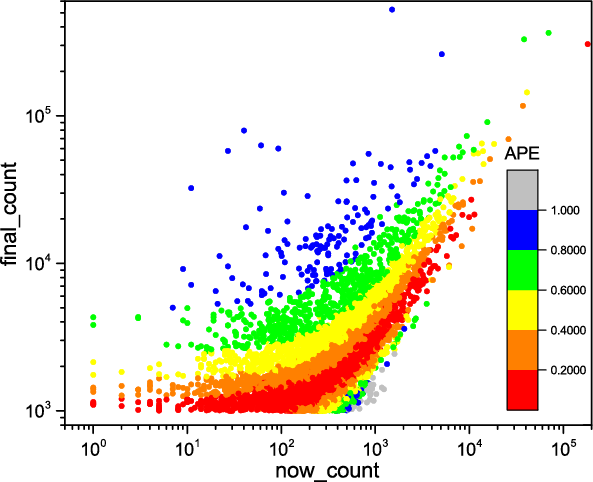}
  &\includegraphics[scale = 0.3]{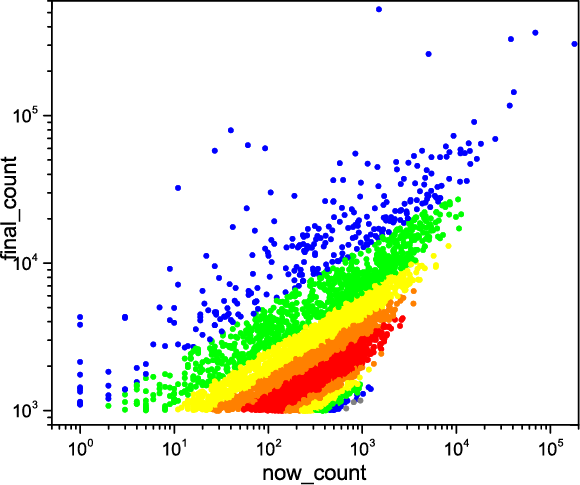}
\\
\includegraphics[scale = 0.4]{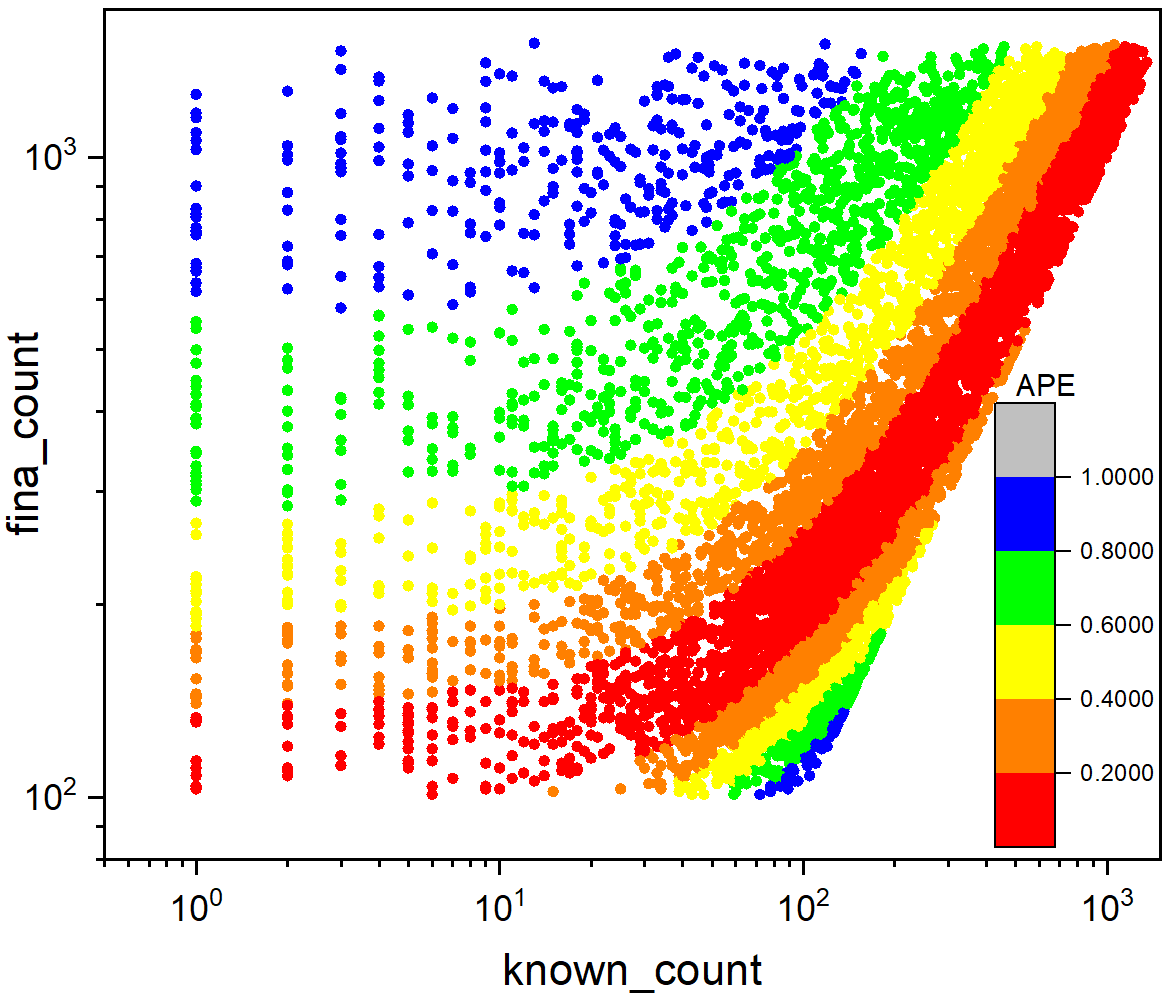}
  &\includegraphics[scale = 0.4]{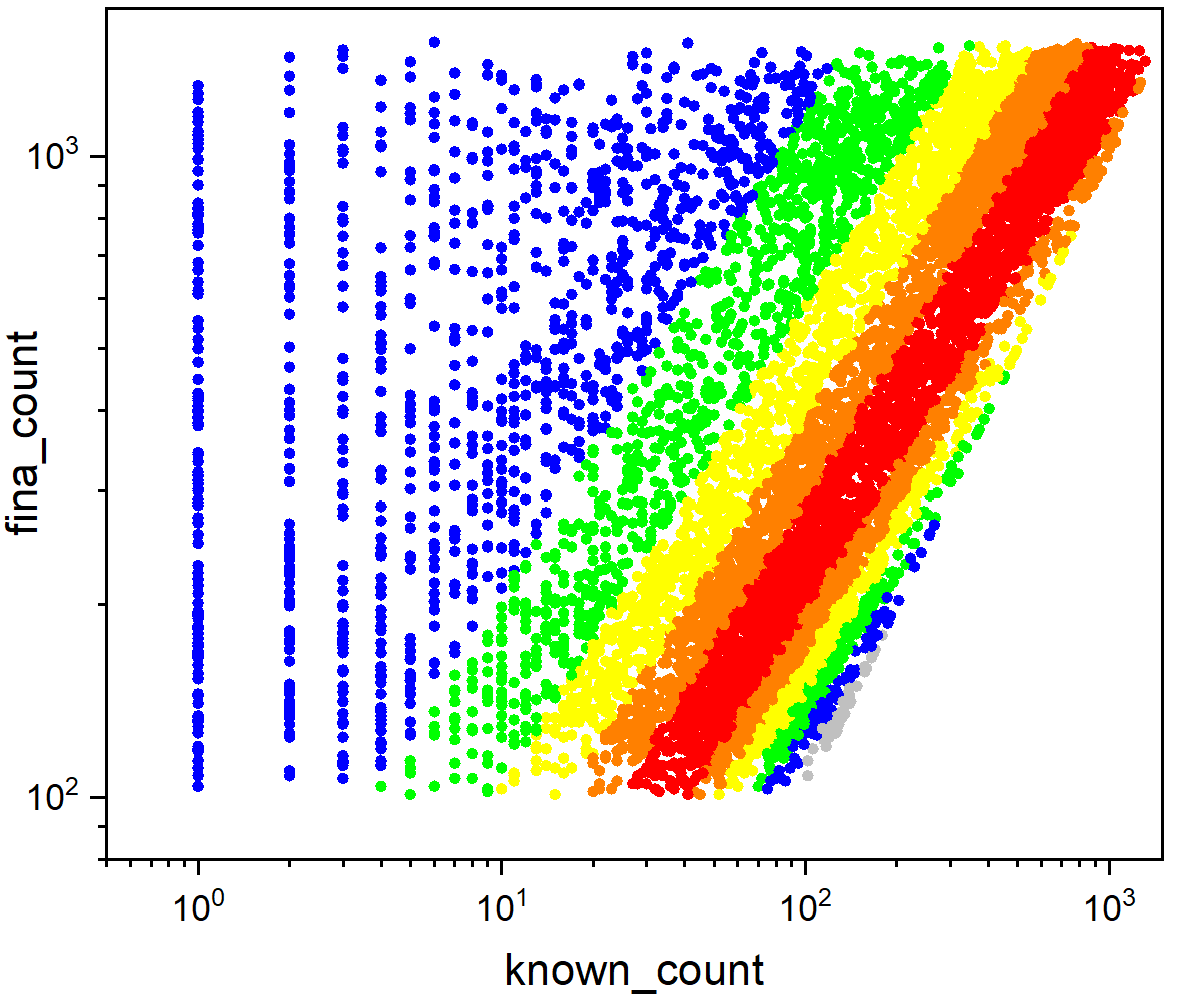} 
\\
  (a) AD algorithm   & (b) Baseline algorithm
\end{tabular}
\caption{APE distribution on utilizing the initial 120-minute data to predict the number of messages forwarded in the next 7 days. The X-axis represents the number of messages forwarded in the first 120 minutes, and the Y-axis represents the total number of messages forwarded in 7 days. The colored bars indicate the size of the APE. The upper part of the figure represents the experimental WeChat data results. The lower part of the figure represents the experimental Weibo data results. }
\label{fig5}
\end{figure*}

In Figure \ref{fig5}, we compare the predictive performance between the AD algorithm and the Baseline algorithm. Obviously, the AD algorithm has a wider range of high prediction accuracy. Intuitively, the red area represents the smallest error (less than 0.2). Compared with the BS algorithm, the AD algorithm can predict the future forwarding amount more accurately (the known forwarding amount ranges from about 1-10000), while the BS algorithm can only reach this standard in the known forwarding amount range (about 50-3000). Whether or not the information is popular in the future, the AD algorithm can give more accurate predictions. This means that the AD algorithm is more flexible and robust, and its prediction performance is less affected by the known information.

\begin{figure*}[htb]
\centering
\begin{tabular}{ccc}
  \includegraphics[scale = 0.28]{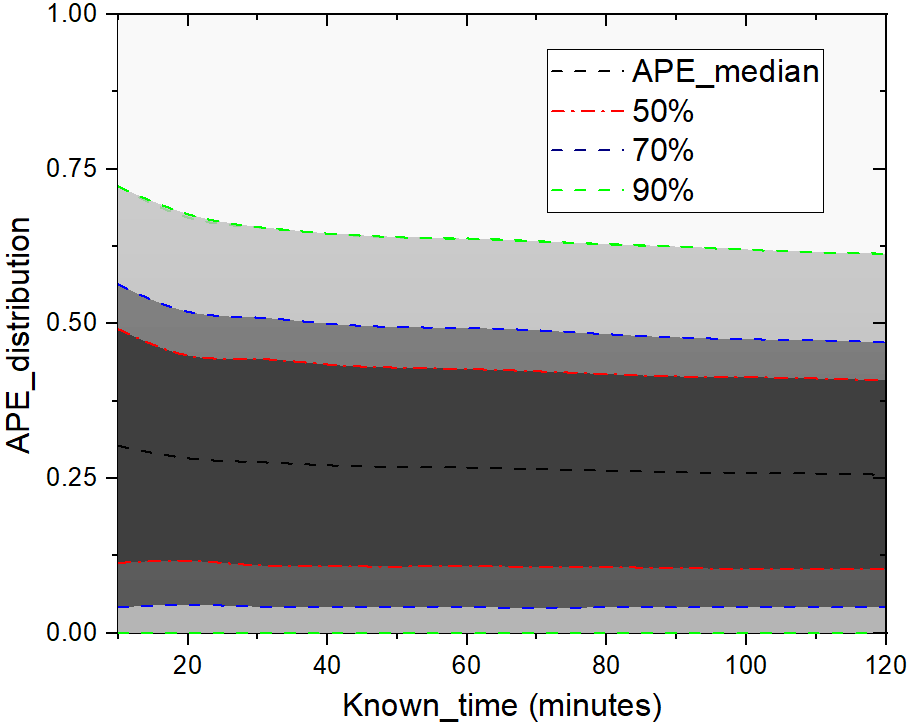}
  &\includegraphics[scale = 0.42]{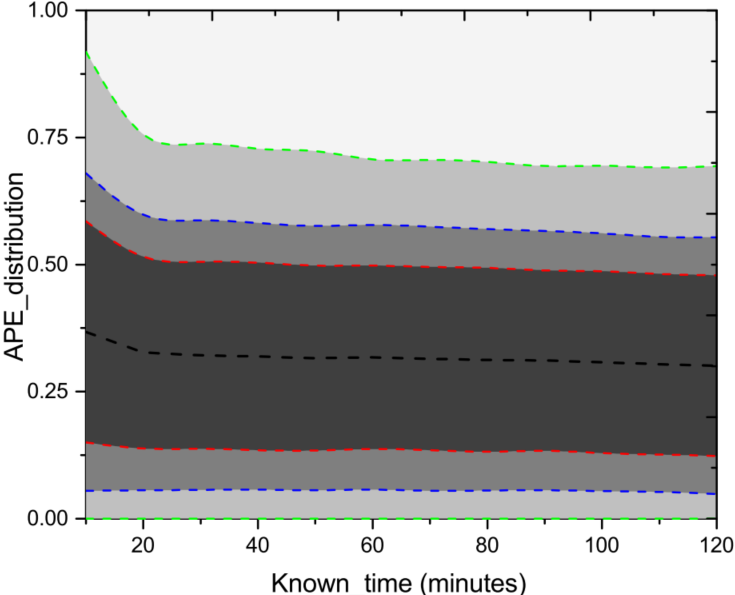}
\\
 \includegraphics[scale = 0.4]{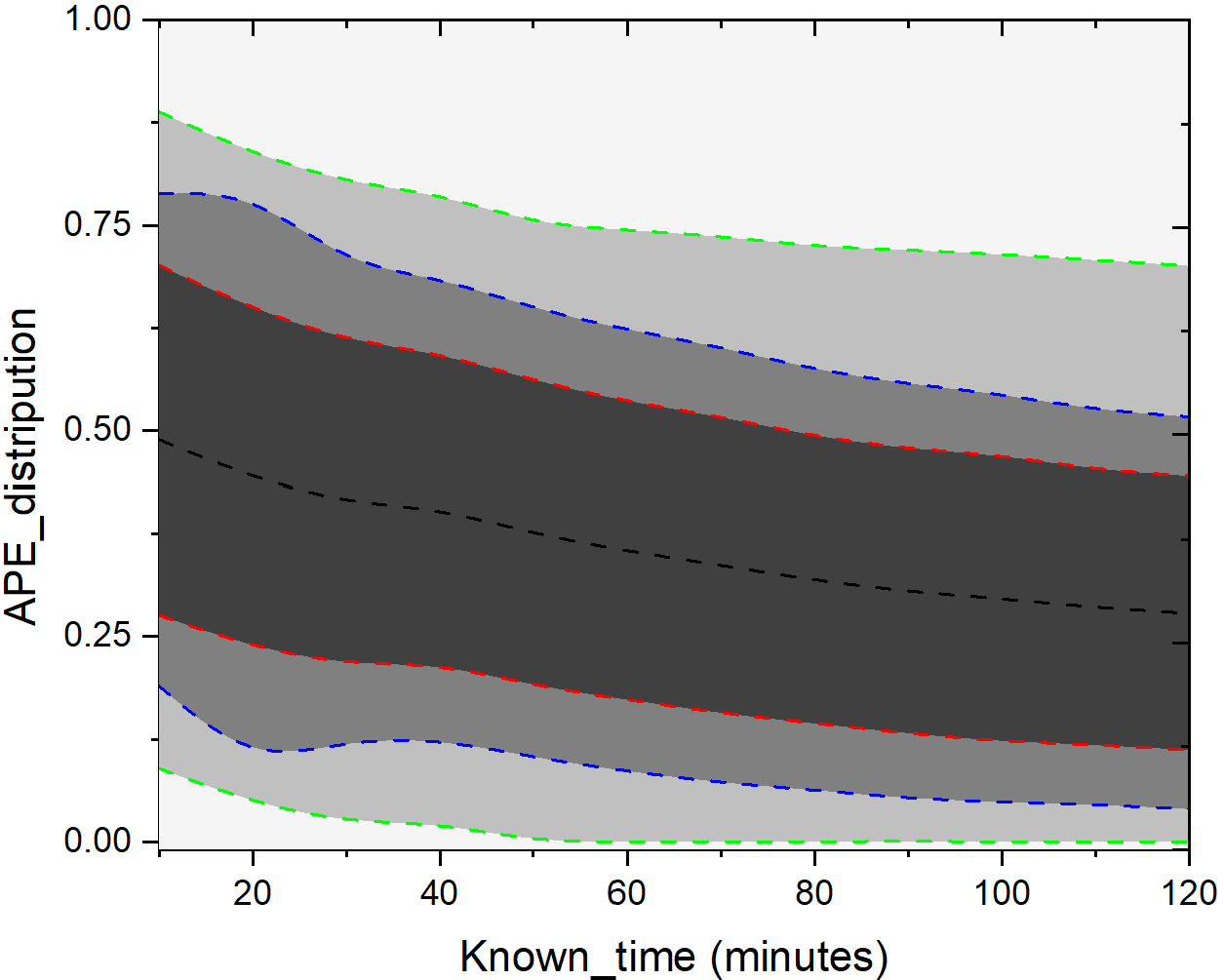}
  &\includegraphics[scale = 0.4]{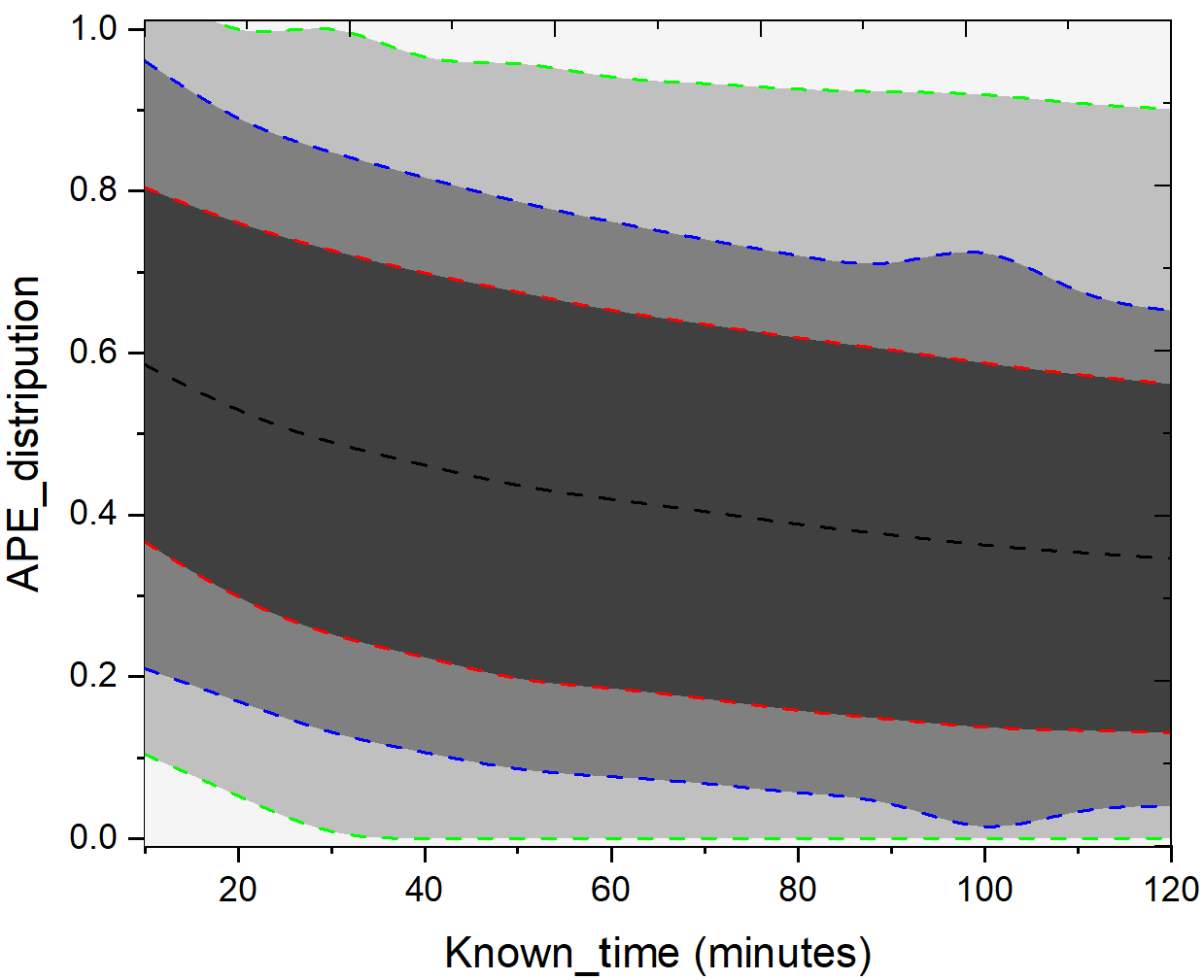}
  \\
 (a) AD algorithm   & (b) Baseline algorithm
\end{tabular}
\caption{Absolute Percentage Error (APE) distribution of the algorithms in the test set. We show the median and the middle 50th, 70th, and 90th percentiles of the distribution of APE across the forward messages. The upper part of the figure represents the experimental WeChat data results. The lower part of the figure represents the experimental Weibo data results.}
\label{fig7}
\end{figure*}

We run AD and BS algorithms on the test set and compute the APE as a function of time. We plot the quantiles of the distribution of APE of the AD algorithm in Figure \ref{fig7}. The AD method demonstrates a clear improvement over the baseline. Take the upper figure (WeChat data) as an example, after 30 minutes, the APE of both algorithms was only in a stable state. After observing the cascade for 20 minutes,  for the AD algorithm, the 90th, 70th, and 50th percentiles of APE are less than $75.6\%, 54.2\%$, and $37.8\%$, respectively. This means that after 20 minutes, the average error is less than $37.8\% $for $50\%$ of the messages and less than $71\%$ for $90\%$ of the messages. After 30 minutes the error gets stable—APE for $90\%$, $70\%$ and $50\%$ of the messages drops to $73.8\%$, $53\%$ and $36.8\%$, respectively. At the same time, the degree of shadow location indicated in the figure indicates that the AD algorithm has greater prediction accuracy than the BS algorithm.

\begin{figure*}[htb]
\centering
\begin{tabular}{ccc}
  \includegraphics[scale = 0.32]{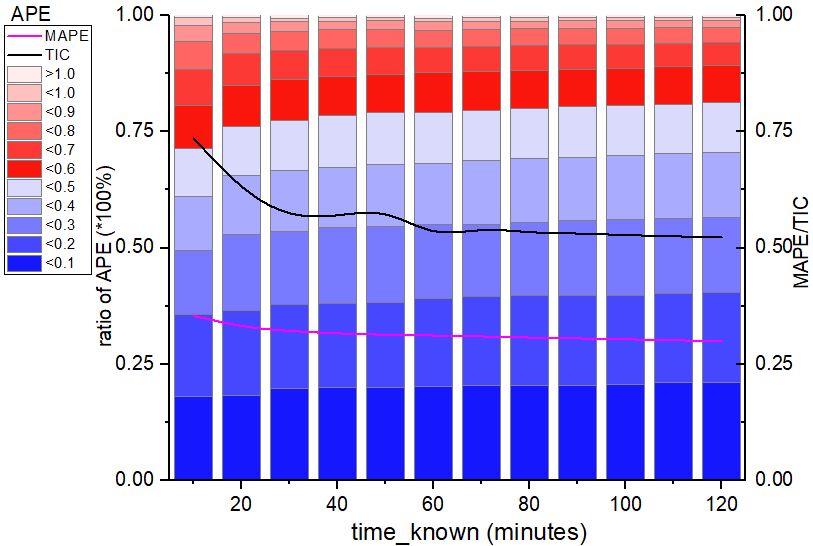}
  &\includegraphics[scale = 0.32]{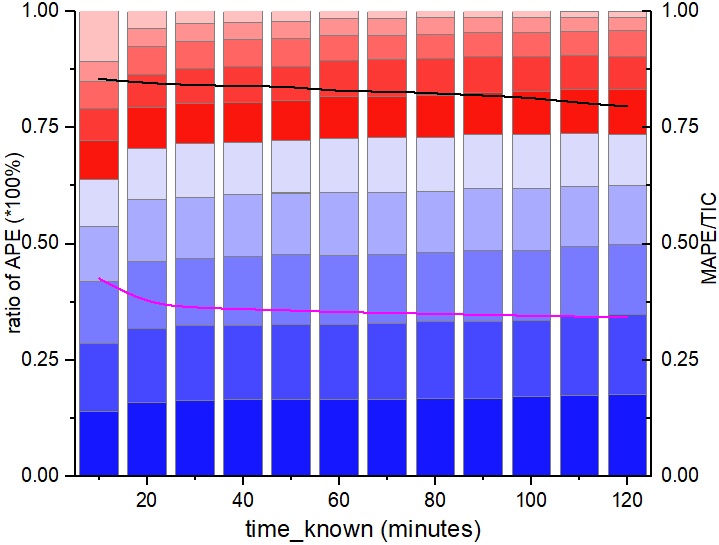}
  \\
    \includegraphics[scale = 0.32]{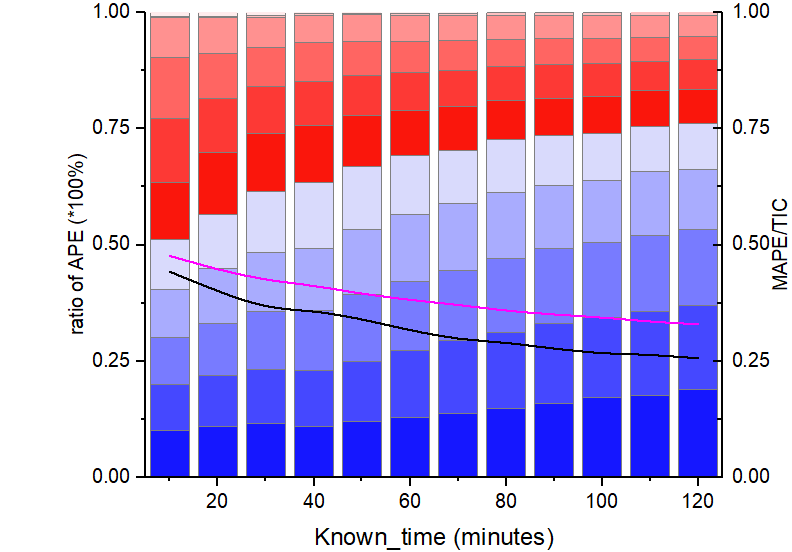}
  &\includegraphics[scale = 0.32]{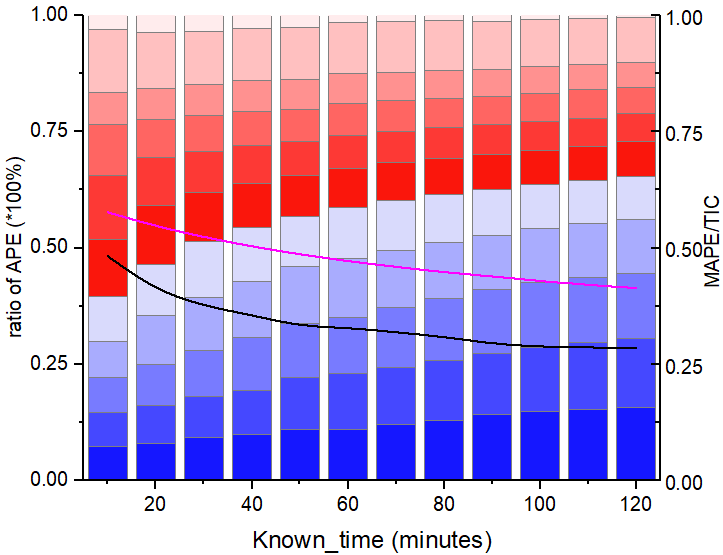}
  \\
  (a) AD algorithm   & (b)Baseline algorithm 
\end{tabular}
\caption{The APE distribution and the MAPE and TIC index varies with knowing $T_{known}$ period of information when predicting the final forward amount after seven days. The X-axis is the time of the known information set, and Y-axis is the ratio of the APE for predicting the final forward amount of messages. Compared with the BS method of predicting the popularity of information, the AD method obviously outperforms in every way. The upper part of the figure represents the experimental WeChat data results. The lower part of the figure represents the experimental Weibo data results.}
\label{fig6}
\end{figure*}
We make a more comprehensive presentation of the errors,  plotting the AMPE, TIC, and the distribution of APE in a graph, and take these error indicators as a function of the known information forwarding time,  as shown in Figure \ref{fig6}. The greater the blue coverage area, the higher the algorithm's prediction impact. Again AD algorithm is giving much more accurate rankings than the baseline algorithm in every way.

\newpage
\subsection{Determine the Peak {$Q_{peak}$}}
In our AD algorithm, there is a very significant variable,$Q_{max}$. During the implementation, we found that if $Q_{max}$ is the peak value $Q_{peak}$ in the process of information forwarding($Q_{max} = Q_{peak}$), the prediction accuracy of AD algorithm will be greatly improved, as shown in Figure \ref{fig9}.
$Q_{peak}$ is the maximum value of the time series of information forwarding volume in the whole life cycle. It is different from $Q_{max}$, which is the maximum value of the time series of information forwarding volume in the known period $T_{known}$. We use the amount of information forwarded in the $T_{known}$ to predict the total amount of information forwarded in the life cycle (7 days in this paper). The experimental results show that whether the $Q_{peak}$ of information occurs within the known time $T_{known}$ will directly affect the prediction accuracy.

\subsubsection{Peak Time $t_{peak}$}
\begin{figure}[H]
  \centering
  \includegraphics[width=10cm]{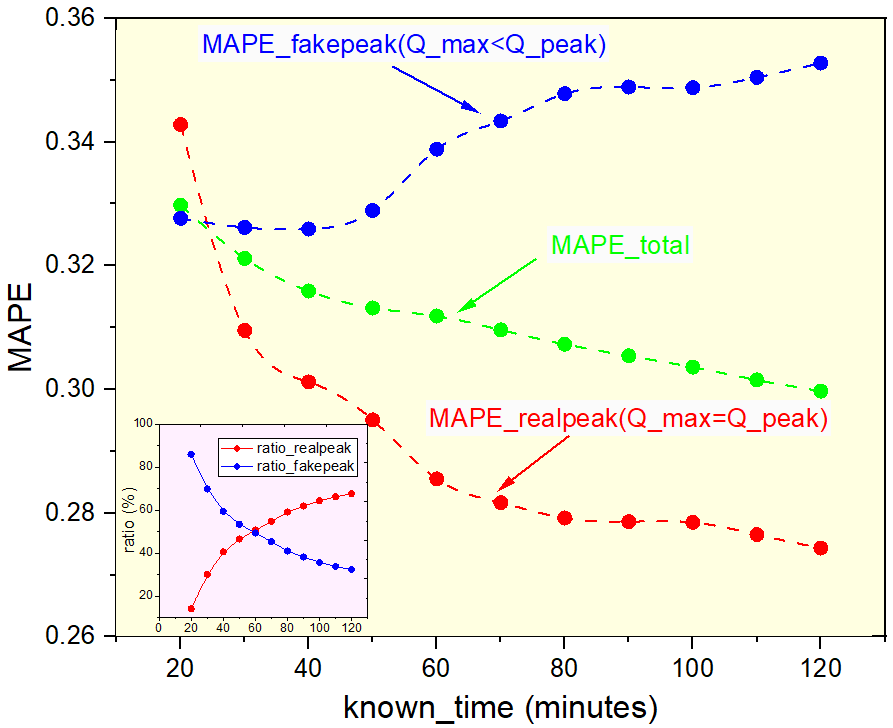}
  \caption{MAPE of the messages varies with the knowing information in the AD algorithm on the WeChat data set. The X-axis is the time of the known information set, and Y-axis is the MAPE for predicting the final forward amount of messages. The red line represents the messages that have got their  $Q_{peak}$ by $T_{known}$, while the blue line means the messages have not got their peak $Q_{peak}$ by $T_{known}$. The internal graph is the ratio of true and fake peaks in information propagation over the first known 120 minutes. AD algorithm can predict more accurately when the $Q_{peak}$ of the message is known.}
  \label{fig9}
\end{figure}

Peak time $t_{peak}$, we refer to the time when the popularity reaches the highest value $Q_{peak}$ per unit time once the popularity evolution starts.  That means we can get $Q_{max} = Q_{peak}$ if $t_{peak}<T_{known}$. The longer the known time $T_{known}$, and the greater the probability of the real peak $Q_{peak}$ appearing, the more accurate the prediction result is.
See Figure \ref{fig9}, MAPE\_realpeak, which signifies that the $Q_{peak}$ value emerged within $t_{peak} < T_{known} = 120$ minutes, i.e.,$Q_{max} = Q_{peak}$, which we term the real peak, as illustrated by the red dot in Figure \ref{fig9}. MAPE\_fakepeak, which indicates $t_{peak} > T_{known} = 120$ minutes, that is, $Q_{peak}$ did not emerge within the known 120 minutes, then $Q_{max} < Q_{peak}$, we use its maximum value $Q_{max}$ to predict, evidently its prediction accuracy rate is obviously lower than $Q_{max} = Q_{peak}$, see the blue dot in Figure \ref{fig9}. The real forecast result is the outcome of combining the aforementioned two conditions, as represented in Figure \ref{fig9}'s green dot schematic design. As a result, the most crucial issue we should examine in our future work is how to determine or forecast $Q_{peak}$. In the first known 120 minutes of message spread data, using the peak $Q_{peak}$ to predict the final counts, the MAPE can reach 0.27, while the fake peak result is 0.35.

\begin{figure}[H]
  \centering
  \includegraphics[width=14cm]{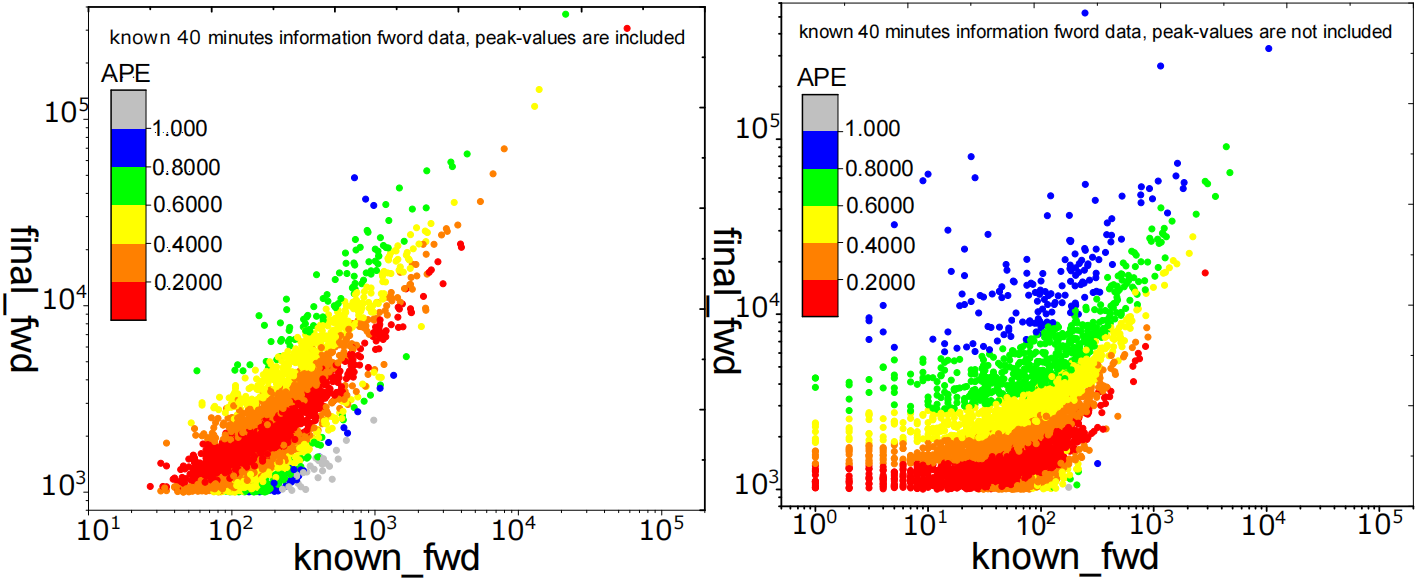}
  \caption{APE distribution of the messages in AD algorithm on the WeChat data set when the peak forward amount $Q_{peak}$ is known (left panels) and not known (right panels). The X-axis represents the number of messages forwarded in the known time $T_{known}$, and the Y-axis represents the total number of messages forwarded in 7 days.}
  \label{fig10}
\end{figure}

To more intuitively assess the impact of $Q_{peak}$ on the prediction outcomes, we partition the data set into two portions for prediction using $t_{peak} < T_{known}$ and $t_{peak} > T_{known}$($T_{known} = 40min$, with WeChat Official Account, it takes less than 30 minutes on average for a message to reach its peak from generation to transmission per unit time, see Figure \ref{fig2}). 
In Figure \ref{fig10}, the peak $Q_{peak}$ has been reached in the left figure, i.e.,$Q_{max} = Q_{peak}(t_{peak} < T_{known})$, and that the colored spots with $APE < 0.4$ account for $ 70.7\%$ of the total. Its final retweets range from $10^3$ to $10^5$(Y axis). Nevertheless, the peak $Q_{peak}$ is not attained in the right figure, i.e.,$Q_{max} < Q_{peak}(t_{peak} > T_{known})$, the colored points with $APE < 0.4$ account for $65.1\%$ of the total, and the final forwarding volume range is only from $10^3$ to $10^4$ (Y axis). This demonstrates that $Q_{peak}$ has a considerable influence on the final forwarding amount range. The determination of the peak $Q_{peak}$ may not only broaden the forecast range of information popularity, but it can also considerably enhance information popularity predictability.

\section{Conclusions}

The spread of information, ideas, innovation, influence, behaviors, and styles within social networks is ubiquitous~\cite{Pastor2015,Dirk2013}. The popularity prediction of information on social platforms is a hot research topic recently \cite{zhang2022anytime,chen2022graph}. Nonetheless, the majority of current methodologies either heavily depend on intricate features that are time-dependent and arduous to extract from multilingual and cross-platform content, or rely on intricate network structures or properties that are frequently challenging to acquire.
In this paper, 
we analyzed several empirical data sets and found that the information-cascading process is best characterized as an activate-decay dynamical process. Then, we introduced the Activate-Decay-based (AD) algorithm, which predicts the long-term forwarding amount of information without requiring knowledge of social network structure or content features. Instead, the AD algorithm only uses limited information, that is, the amount of information forwarded within specific time intervals (e.g., 30 minutes for WeChat, and 3 minutes for Weibo), to predict the total forwarding amount over several days accurately.

The AD algorithm is a straightforward and practical approach for predicting information popularity, which outperforms the baseline algorithm in accuracy. However, a challenge remains in determining the actual maximum forwarding amount within a given time interval. To address this challenge, we assume that the maximum propagation amount per unit time based on past data denoted as $Q_{max}^{real}$, represents the peak value. Nonetheless, we find that identifying the genuine peak forwarding value can further improve the accuracy of our prediction results, as illustrated in Figure \ref{fig9}. Therefore, we plan to focus on this issue in future research.

\vspace{6pt} 

\section*{Author Contributions}
{Methodology, L.W., and X.-L.R.; formal analysis, L.W., L.Y., X.-L.R., and L.L.; resources, L.Y. and L.L.;  writing---original draft preparation, L.W., and X.-L.R.; writing---review and editing, L.W., L.Y., X.-L.R., and L.L.; visualization, L.W.; supervision, X.-L.R, and L.L.; funding acquisition, L.L., and X.-L.R. All authors have read and agreed to the published version of the manuscript.}

\section*{Funding}
{This research was funded by the STI 2030--Major Projects (2022ZD0211400), the China Postdoctoral Science Foundation (2022M710620), the Sichuan Science and Technology Program (2023NSFSC1353), the Project of Huzhou Science and Technology Bureau (2021YZ12), and  the UESTCYDRI research start-up (U032200117). This work has been partially supported by the New Cornerstone Science Foundation through XPLORER PRIZE.}

\section*{Data Availability}
{ The data from Weibo in this study will be available at the following GitHub repository: \url{https://github.com/renxiaolong/InformationPopularityPrediction} after this paper is accepted. The data set of WeChat was generated during a collaboration project with Tecent’s WeChat department. All the WeChat data is kept within the company.} 

\section*{Conflict of Interest}
\textbf{The authors declare no conflict of interest.} 



\bibliographystyle{unsrt}
\bibliography{main}


\end{document}